\documentclass[aps,prl,a4paper,twocolumn,showpacs,superscriptaddress,floatfix]{revtex4}

\usepackage{graphicx,amsmath,bbm,mathrsfs,amssymb,pstricks,times,psfrag}
\begin{document}
\title{Full characterization of Gaussian bipartite entangled states by a
single homodyne detector}
\author{V.~D'Auria}
\affiliation{CRS Coherentia CNR-INFM, Napoli, Italia.}
\author{S.~Fornaro}
\affiliation{Dipartimento di Scienze Fisiche Universit\`a ``Federico II'',
Napoli, Italia.}
\author{A.~Porzio}
\affiliation{CRS Coherentia CNR-INFM, Napoli, Italia.}
\affiliation{CNISM UdR Napoli Universit\`a, Napoli, Italia.}
\author{S.~Solimeno}
\affiliation{Dipartimento di Scienze Fisiche Universit\`a ``Federico II'',
Napoli, Italia.}
\affiliation{CNISM UdR Napoli Universit\`a, Napoli, Italia.}
\author{S.~Olivares}
\affiliation{CNISM UdR Milano Universit\`a, Milano, Italia.}
\affiliation{Dipartimento di Fisica dell'Universit\`a di Milano,
Milano, Italia.}
\author{M.~G.~A.~Paris} 
\affiliation{Dipartimento di Fisica dell'Universit\`a di Milano,
Milano, Italia.}
\affiliation{CNISM UdR Milano Universit\`a, Milano, Italia.}
\affiliation{ISI Foundation, Torino, Italia.}
%%%%%%%%%%%%%%%%%%%%%%%%%%%%%%%%%%%%%%%%%%%%%%%%
\begin{abstract}
We present the full experimental reconstruction of Gaussian entangled
states generated by a type--II optical parametric oscillator (OPO) below
threshold. Our scheme provides the entire covariance matrix using a
single homodyne detector and allows for the complete characterization of
bipartite Gaussian states, including the evaluation of purity,
entanglement and nonclassical photon correlations, without a priori
assumptions on the state under investigation.  Our results show that
single homodyne schemes are convenient and robust setups for the full
characterization of OPO signals and represent a tool for quantum
technology based on continuous variable entanglement.  
\end{abstract}
\date{\today}
\maketitle
%%%%%%%%%%%%
\indent {\em Introduction}---In this letter we address the complete 
experimental characterization of bipartite Gaussian entangled states.
In our experiment continuous-wave (CW) entangled light beams are
generated by a single type--II optical parametric oscillator (OPO) below
threshold, and then their covariance matrix (CM) is fully reconstructed
using a novel scheme \cite{sh05} that involves a single homodyne detector
\cite{ray99}.
To our knowledge this is the first {\em complete} characterization of
OPO signals without a priori assumptions and paves the way to a deeper
investigation of continuous variable entanglement without experimental
loopholes.
\par
Light beams endowed with nonclassical correlations \cite{eis03} are
crucial resources for quantum technology and find applications in
quantum communication \cite{vnl02}, imaging \cite{lug02} and
precision measurement \cite{dar01,dau05}. Their full characterization has
a fundamental interest in its own and represents a tool for the design of
quantum information processing protocols in realistic conditions.
Remarkably, entangled states produced by OPOs are Gaussian states
\cite{simXX,marYY} and thus may be fully characterized by the first two
statistical moments of the field modes.  In turn, the CM contains the
complete information about entanglement \cite{sim00,dua00}, {\em i.e.}
about their performances as a resource for quantum technology.
\par
Bipartite entangled states may be generated by mixing at a beam splitter
(BS) two squeezed beams obtained by a degenerate OPO below threshold
\cite{bow03}.  The beams exiting the BS are entangled \cite{theZZ} and a
partial reconstruction of the corresponding CM has been obtained
% by measuring the noise relative to the sum/difference variances of the modes
\cite{bow04}. The complete reconstruction of a CM has been obtained for
different entangled states, by varying single-mode squeezing
\cite{dig07}.  In this configuration two OPOs are used and the
amount of entanglement critically depends on the symmetry between the two
squeezed beams. From the experimental point of view this requires an
accurate setting on the two squeezers and a strict control on the
relative phase.
The measured CM presents some unexpected deviations from a proper
form so leaving a question open on the reliability of double homodyne 
schemes due to technical difficulties \cite{bow04,lau05}.
A more direct way to generate quadrature entanglement
is to use a single non-degenerate OPO \cite{dru90} which represents a
robust and reliable source of EPR-type correlation either below
\cite{zha99,kim92} or above threshold \cite{jin06,bre07,kel08}. 
Partial reconstructions of the CM in the pulsed regime has been achieved for
the spectrally degenerate but spatially non-degenerate twin beams at the
output of a type--I parametric amplifier \cite{wen04}, whereas in the CW
regime cross polarized beams emitted by a self-locked type--II OPO have
been examined at frequency degeneracy \cite{lau05}.
Although the correlation properties of OPO signals have been widely
investigated, no proper CM reconstruction has been performed so
far. In turn, in previous proposals and experiments non-physical CM
entries \cite{bow03,bow04,wen04,lau05} or deviations from a proper CM
\cite{dig07} appeared, thus requiring a priori hypothesis on  the measured
state to understand the experimental results.
\par
In this letter we report the first complete measurement of the CM for
the output of a single non-degenerate OPO. The two entangled beams are
emitted with orthogonal polarization and degenerate frequency by a CW
type--II OPO below threshold. In order to reconstruct the ten
independent elements of the CM, the beams are optically combined into
six auxiliary modes, whose quadratures are measured using a single
homodyne detector \cite{sh05}. The first two moments of the relevant
quadratures are obtained by tomographic reconstruction using the whole
homodyne data set, the CM is, then, fully reconstructed after assessing
the Gaussian character of the signal and compared with a general model
describing a realistic OPO. Entanglement is demonstrated using the
partial transpose method \cite{sim00}, the Duan inequality \cite{dua00}
and the stricter EPR criterion \cite{tre05}, and quantified upon
evaluating the logarithmic negativity and the entanglement of formation
(EoF). We also reconstruct the joint
photon number distribution and demonstrate nonclassical photon
correlations by evaluating the noise reduction factor.
%Our experimental
%setup makes use of a single OPO and a single homodyne detector and thus
%represent a compact and robust setup for entanglement generation and
%characterization.
In the following, after defining notation and a
brief summary of the reconstruction method, we describe in details the
apparatus and the experimental results.
%%%%%%%%%%%%%%%%%%
\par{\em Reconstruction method---}Upon introducing the vector
$\boldsymbol{R}=(x_{1},y_{1},x_{2},y_{2})$ of canonical operators, in
terms of the mode operators $a_{k}$,  $x_{k}=(a_{k}^{\dag
}+a_{k})/\sqrt{2}$, $y_{k}=i(a_{k}^{\dag }-a_{k})/ \sqrt{2}$, $k=1,2$,
the CM $\boldsymbol{\sigma}$ of a bipartite state $\varrho$ is defined
as the block matrix
$\boldsymbol{\sigma }=\left( \begin{array}{c|c} A & C \\ \hline
C^{T} & B \end{array} \right)$,
$\sigma _{hk}=\frac{1}{2}\langle \{R_{k},R_{h}\}\rangle -\langle
R_{k}\rangle \langle R_{h}\rangle$ 
where $A$, $B$ and $C$ are $2\times 2$ real matrices, $\langle O\rangle
= \mathrm{Tr}(\varrho \,O)$ and $\{f,g\}=fg+gf$. In the following we
will use the notation $a\equiv a_{1}$ and $b\equiv a_{2}$ and also
consider the four additional auxiliary modes  $c=(a+b)/\sqrt{2}$,
$d=(a-b)/\sqrt{2}$, $e=(ia+b)/\sqrt{2}$, and $f=(ia-b)/\sqrt{2}$
obtained by the action of polarizing beam splitters (PBS) and
phase-shifters on modes $a$ and $b$.  Positivity of the density matrix
for physical states is written in terms of the uncertainty relation for
the minimum symplectic eigenvalue $\nu _{-}$ of the CM, {\em i.e.}
$\nu_{-}\geq 1/2$.
For Gaussian states, the state purity is given by $\mu (\boldsymbol{
\sigma })=(4\sqrt{\mathrm{Det[\boldsymbol{\sigma }]}})^{-1}$ whereas
separability corresponds to positivity of the partially transpose (PPT)
density matrix. A bipartite Gaussian state is separable iff
$\tilde{\nu}_{-}>1/2 $, where $\tilde{\nu}_{-}$ is the minimum
symplectic eigenvalue of $\boldsymbol{\Delta } \boldsymbol{\sigma
}\boldsymbol{\Delta }$, $ \boldsymbol{\Delta }=
\mathrm{Diag}[1,1,1,-1]$. A convenient measure of entanglement is thus
given by the logarithmic negativity $E_{\mathcal{N}}(\boldsymbol{ \sigma
})=\mathrm{max}(0,-\ln 2 \tilde{\nu}_{-})$ \cite{vid02} and
the EoF $E_{\cal F}(\boldsymbol{ \sigma})$ can be evaluated
following Ref.~\cite{MM:08}. In addition
to $E_{\mathcal{N}}(\boldsymbol{\sigma })$ the Duan criterion gives a
necessary condition for non-separability in terms of the noise
properties of $c$ and $d$ \cite{dua00} whereas a stricter condition
\cite{tre05}, referred to as EPR criterion, involves the conditional
variances on $x_{a}$ and $y_{a}$ obtained from a measurement of $x_{b}$
and $y_{b}$ for the explicit expressions in terms of the noise on modes
$a$, $b$, $c$ and $d$. 
\par
In  our experiment, the block $A$ of the CM is retrieved by measuring
the single-mode quadratures of mode $a$: the variances of $x_{a}$ and
$y_{a}$ give the diagonal elements, while the off diagonal ones are
obtained from the additional quadratures $z_{a}\equiv \left(
x_{a}+y_{a}\right) /\sqrt{2}$ and $\ t_{a}\equiv \left(
x_{a}-y_{a}\right) /\sqrt{2}$ as $\sigma _{12}= \sigma
_{21}=\frac12(\langle z_{a}^{2}\rangle -\langle t_{a}^{2}\rangle
)-\langle x_{a}\rangle \langle y_{a}\rangle$ \cite{sh05}. The block $B$
is reconstructed in the same way from the quadratures of $b$, whereas
the elements of the block $C$ are obtained from the quadratures of the
auxiliary modes $c$, $d$, $e$ and $f$ as follows
$\sigma _{13}= \frac12(\langle x_{c}^{2}\rangle -\langle x_{d}^{2}\rangle
)-\langle x_{a}\rangle \langle x_{b}\rangle$, 
$\sigma _{14}= \frac12 (\langle y_{e}^{2}\rangle -\langle y_{f}^{2}\rangle
)-\langle x_{a}\rangle \langle y_{b}\rangle$, 
$\sigma _{23}= \frac12 (\langle x_{f}^{2}\rangle -\langle x_{e}^{2}\rangle
)-\langle y_{a}\rangle \langle x_{b}\rangle$,
$\sigma _{24}= \frac12 (\langle y_{c}^{2}\rangle -\langle y_{d}^{2}\rangle
)-\langle y_{a}\rangle \langle y_{b}\rangle$.
Notice that the measurement of the $f$-quadratures is not mandatory, 
since $\langle x_{f}^{2}\rangle =\langle x_{b}^{2}\rangle
+\langle y_{a}\rangle ^{2}-\langle x_{e}^{2}\rangle $ and $\langle
y_{f}^{2}\rangle =\langle x_{a}^{2}\rangle +\langle y_{b}^{2}\rangle
-\langle y_{e}^{2}\rangle $. Analogous expressions hold for $\langle
x_{e}^{2}\rangle $ and $\langle y_{e}^{2}\rangle $.
\par
In the ideal case the OPO output is in a twin-beam state
$\mathbf{S}(\zeta )|0\rangle$, ${\mathbf S}(\zeta) = 
\exp\{\zeta a^\dag b^\dag - \bar\zeta ab\}$ being the entangling two-mode 
squeezing operator: the corresponding CM has diagonal blocks 
$A$, $B$, $C$ with the two diagonal elements of
each block equal in absolute value. In realistic OPOs, cavity and crystal 
losses lead to a mixed state, {\em i.e.} to an effective thermal
contribution. In addition, spurious nonlinear processes, not perfectly
suppressed by the phase matching, may combine to the down conversion, 
contributing with local squeezings. Finally, due to small misalignments of
the nonlinear crystal, a residual component of the field polarized along $a$
may project onto the orthogonal polarization (say along $b$), thus 
leading to a mixing among the modes \cite{dau08}. Overall,
the state at the output  is expected to be a zero amplitude Gaussian
entangled state, whose general form may be written as
$\varrho_{g} = {\mathbf U} (\beta) {\mathbf S}(\zeta)\, 
{\mathbf{LS}}(\xi_1,\xi_2)\, {\mathbf T}\, 
{\mathbf{LS}}^\dag(\xi_1,\xi_2)\, {\mathbf S}^\dag (\zeta) 
{\mathbf U}^\dag (\beta)$, where
${\mathbf T}= \tau_1 \otimes \tau_2$, with 
$\tau_k = (1+ \bar n_k)^{-1} [\bar n_k/(1+\bar n_k )]^{a^\dag a}$ 
denotes a two-mode thermal state with $\bar n_k$ average photons
per mode,  ${\mathbf{LS}}(\xi_1,\xi_2)= S(\xi_1) \otimes S(\xi_2)$,
$S(\xi_k)=\exp\{\frac12 (\xi_k a^{\dag 2} - \bar\xi_k a^2)\}$ denotes
local squeezing and ${\mathbf U}(\beta)= \exp\{\beta a^\dag b 
- \bar\beta ab^\dag\}$ a mixing operator, $\zeta$, $\xi_k$ and $\beta$
being complex numbers. 
For our configuration, besides a thermal contribution due to internal
and coupling losses, we expect a relevant entangling contribution
with a small residual local squeezing and, as mentioned above, a possible 
mixing among the modes. The CM matrix corresponding to $\varrho _{g}$ has 
diagonal blocks $A$, $B$, and $C$ with possible
asymmetries among the diagonal elements.
\par
\begin{figure}[h]
\includegraphics[width=0.45\textwidth]{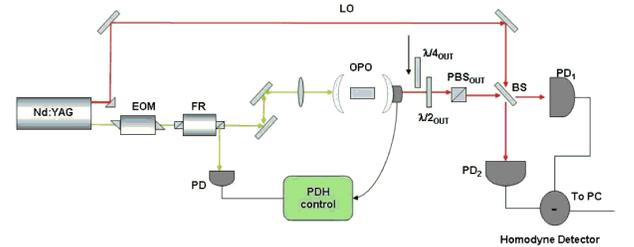}
\vspace{-0.3cm}
\caption{
Experimental setup: A type-II OPO containing a
periodically poled crystal (PPKTP) is pumped by the second harmonic of a
Nd:YAG laser. At the OPO output, a half-wave plate
($\lambda /2_{\rm out}$), a quarter-wave plate ($\lambda /4_{\rm out}$) and a
PBS$_{\rm out}$ select the mode for homodyning.
The resulting electronic signal is acquired via a PC module.} \label{f:setup}
\end{figure}

{\em Experimental setup}---The experimental setup, shown in Fig.~\ref{f:setup},
relies on a CW internally frequency doubled Nd:YAG laser
pumping (@532~nm) a non degenerate OPO based
on a periodically poled $\alpha $--cut KTP (PPKTP) crystal
(\textit{Raicol Crystals Ltd}. on custom design) \cite{fei92}.
The use of the $\alpha $-cut
PPKTP allows implementing a type-II phase matching with cross polarized
signal ($a$) and idler ($b$) waves, frequency degenerate @1064~nm for a
crystal temperature of $\approx 53^{\circ }$C.
The OPO cavity is locked to the pump beam by Pound-Drever technique
\cite{dre83} and adjusted to work in triple resonance by finely tuning its
geometrical properties \cite{dau08}. The cavity output coupling @1064~nm is $%
\approx 0.73$, corresponding to an experimental line-width of $16$ MHz
@1064~nm. The measured oscillation threshold is $P_{th}\approx 50$ mW; during
the acquisition the system has been operated below threshold at 60\% of the
threshold power.
%%%%%%%%%%%%%%%
 \\
%%%%%%%%%%%%%%%
In order to select mode $a$ and $b$ or their combinations $c$ and $d$,
the OPO beams are sent to a half-wave plate and a PBS.
Modes $e$ and $f$ are obtained by inserting an additional
quarter-wave plate \cite{sh05}. The PBS output
goes to a homodyne detector, described in details in \cite{dau05,opt05},
exploiting the laser output @1064~nm as local oscillator (LO).
%The homodyne photocurrent is formed from the output of two
%photodiodes \emph{Epitaxx ETX300}, both matched to a low--noise
%trans--impedance AC ($>$ few kHz) amplifier. The difference-photocurrent is
%further amplified by a low noise high gain amplifier \emph{Miteq AU1442}.
The overall homodyne detection efficiency is $\eta =0.88 \pm 0.02$. The LO
reflects on a piezo-mounted mirror (PZT), which allows varying its phase $%
\theta $. In order to avoid the laser low frequency noise, data sampling is
moved away from the optical carrier frequency by mixing the homodyne current
with sinusoidal signal of frequency $\Omega =3$~MHz \cite{opt05}. The
resulting current is low--pass filtered ($B=300$~kHz) and sampled by a PCI
acquisition board (Gage 14100, 1M--points per run, 14 bits resolution). The
total electronic noise power has been measured to be $16$~dBm below the
shot--noise level, corresponding to a signal to noise ratio of about $40$. 
%%%%
\par{\em Reconstruction and experimental results}---Acquisition is triggered
by a linear ramp applied to the PZT and adjusted to obtain a $2\pi $
variation in $200$ ms. Upon spanning the LO phase $\theta $,
the quadratures $x\left( \theta \right) =x\cos \theta +y\sin \theta $
are measured. Calibration with respect to the noise of the vacuum state
is obtained by acquiring a set of data with the output from the OPO
obscured. All the expectation values needed to reconstruct ${\boldsymbol
\sigma }$ are obtained by quantum tomography \cite{qht03}, which
allows to compensate nonunit quantum efficiency and to reconstruct any
expectation value, including those of specific quadratures and their
variances, by averaging special pattern functions over the whole data set.
As a preliminary check of the procedure, we verified
that the CM of the vacuum state is consistent with
${\boldsymbol\sigma}_0 = \frac12\,\mathbb{I}$ within the experimental
errors.  
%%%%%%%%%%%%%
\begin{figure}[h]
\includegraphics[width=0.4\textwidth]{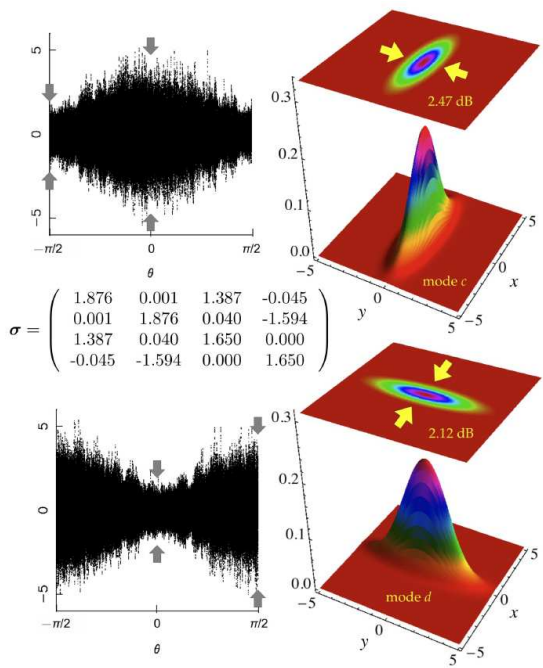}
\vspace{-0.3cm}
\caption{(Left): experimental homodyne traces and reconstructed CM;
(Right): reconstructed Wigner functions. Top plots are for modes $c$ 
and bottom ones for $d$. Arrows on the 
homodyne plots show the positions of the maximum and minimum variances. 
$\theta$ is the relative phase between the signal and the LO.
\label{f:wig:cd}}
\end{figure}
%%%%%%%%%%%%%
\par
We start our analysis by checking the Gaussian character of the OPO
signals, {\em i.e.} upon evaluating the Kurtosis of homodyne
distribution at fixed phase of the LO \cite{opt05}.
Besides, we checked that the mean
values of all the involved quadratures are negligible, in agreement with
the description of OPO output as a zero amplitude state.  Then, we have
measured the  quadratures of the six modes $a$-$f$.  We found the modes
$a$ and $b$ excited in a thermal state, thus confirming the absence of
relevant local squeezing.  Their combinations $c$, $d$, $e$ and $f$ are
squeezed thermal states with squeezing appearing on $y_{c}$,
$x_{d}\,$,$\ t_{e}$ and $z_{f}$, respectively.  In Fig.~\ref{f:wig:cd}
we show the experimental homodyne traces for modes $c$ and $d$ as well
as the corresponding Wigner functions, obtained by reconstructing the
single-mode CM.  As it is apparent from the plots both modes are
squeezed with quadratures noise reduction, corrected for nonunit
efficiency, of about $2.5$~dB. An analogue behavior has been observed
for modes $e$ and $f$.  The CM of Fig. \ref{f:wig:cd} indeed reproduce
that of an entangled thermal state with small corrections due to local
squeezing and mixing.  The relevant parameters to characterize the
corresponding density matrix $\varrho_g$ are the mean number of thermal
photons $\bar n_1 \simeq 0.67$, $\bar n_2 \simeq 0.18$ and entangling
photons $\bar n_s = 2 \sinh^2 |\zeta|\simeq 0.87$ \cite{pars}.  The
errors on the CM elements for the blocks $A$ and $B$ are of the order
$\delta\sigma_{jk} \simeq 0.004$ and have been obtained by propagating
the tomographic errors. In this case phase fluctuations are irrelevant,
since the two modes are both excited in a thermal state. On the other
hand, in evaluating the errors on the elements of the block $C$ the
phase-dependent noise properties of the involved modes have to be taken
into account, and the tomographic error has to be compared with the
error due to the finite accuracy in setting the LO phase $\theta$. The
elements $\sigma _{13}$ and $\sigma _{24}$ are obtained as combinations
of squeezed/anti--squeezed variances, which are quite insensitive to
fluctuations of $\theta $. As a consequence the errors on these elements
are given by the overall tomographic error $\delta\sigma_{jk} \simeq
0.004$. On the other hand, the elements $\sigma _{14}$ and $\sigma
_{23}$ depend on the determination of $x_{e,f}^{2}$ and $y_{e,f}^{2}$,
which are sensible to phase fluctuations. In order to take into account
this effect we evaluate errors as the fluctuations in the
tomographically reconstructed quadratures induced by a $\delta \theta
\simeq 20$~mrad variation in the LO phase, corresponding to the
experimental phase stability of the homodyne detection. The resulting
errors are about $\delta\sigma_{14} = \delta\sigma_{23}\simeq 0.03$ for
both CM elements.  The off-diagonal elements of the three matrices $A$,
$B$ and $C$ are thus zero within their statistical errors, in agreement
with the expectation for an entangled thermal state.  As mentioned
above, the experimental procedure may be somehow simplified exploiting
the relationships among modes, and expressing mode $e$ or $f$ in terms
of the others: only five modes are then needed.
Upon rewriting the off-diagonal terms of $C$ in terms of the five modes
we arrive at $\sigma _{14}=0.02\pm 0.03$ and $\sigma_{23}=0.04\pm 0.03$
when eliminating the mode $f$ and $\sigma_{14}=0.06\pm 0.03$ and
$\sigma_{23}=0.06\pm 0.03$ when eliminating the mode $e$. Both
procedures provide results in agreement with those obtained by using the
complete set of homodyne data for the six modes.
\par
Since the minimum symplectic eigenvalue of ${\boldsymbol\sigma }$ is
$\nu_{-}=0.68\pm 0.02\geq 0.5$, the CM corresponds to a physical state. 
State purity is $\mu (\boldsymbol\sigma) = 
0.31\pm 0.01$. The minimum symplectic eigenvalue for the partial 
transpose is $\tilde{\nu}_{-}=0.24\pm 0.02$, which corresponds to a 
logarithmic negativity $E_{\mathcal{N}}(\boldsymbol{\sigma }%
)=0.73\pm 0.02$, {\em i.e.} the state is entangled, with EoF
$E_{\mathcal{F}}(\boldsymbol{\sigma })=1.46\pm 0.02$.
In turn, it satisfies the Duan inequality with the results $0.29\pm
0.01<1/2$ and the EPR criterion with $0.21\pm 0.01<1/4$.
\par
\begin{figure}[h]
\centerline{\includegraphics[width=0.40\textwidth]{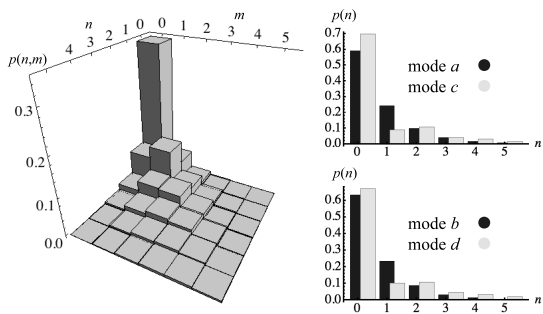}}
\vspace{-0.3cm}
\caption{(Left): Joint photon number distribution $p(n,m)$ for the entangled
state of modes $a$ and $b$ at the output of the OPO. (Right): single-mode 
photon distributions $p(n)$ for modes $a$ and $c$ (top right) and $b$ and
$d$ (bottom right). The single-mode distributions of modes $a$ and $b$
are thermal and correspond to the marginals of $p(n,m)$.
The distributions for modes $c$ and $d$ are those of squeezed thermal
states. \label{f:pnm}} 
\end{figure}

Entangled Gaussian states as $\varrho_{g}$ may be endowed with
nonclassical photon number correlations, {\em i.e} squeezing in the
difference photon number. This may be checked upon evaluating the noise
reduction factor ${\cal R} = \hbox{Var}(D_{ab})/(\bar N_a+\bar N_b)$
where $\hbox{Var} (D_{ab})$ denotes the variance of the difference
photocurrent $D_{ab}= N_a - N_b$, $N=a^\dag a$ being the number
operator, and $\bar N_k = \langle N_k \rangle$ the average
photon number. A value ${\cal R}<1$ is a marker of nonclassical
correlations between the two modes. We obtained ${\cal R}=0.50\pm 0.02$,
in agreement with the theoretical description \cite{deg07} for the
values of thermal and entangling photons reported above. Starting from
the CM one can reconstruct the full joint photon distribution
$p(n,m)$ of the modes $a$ and $b$: the result is shown in Fig.~\ref{f:pnm}
where the correlations between the two modes are clearly seen. We have
also evaluated the single-mode photon distributions (either from
data or from the single-mode CM) for modes $a$-$d$. Results are
reported in Fig.~\ref{f:pnm}: distributions of $a$ and $b$ are thermal,
whereas the statistics of modes $c$ and $d$ correctly reproduces the
even-odd oscillations expected for squeezed thermal states. 
%%%%%
%%%%%
%%%%%%%%%%%%%%%%%%%%%%%%
\par{\em Conclusion}---We have presented the complete
reconstruction of the CM for the output of a CW type II non-degenerate
OPO, below threshold and frequency degenerate.  The CM elements have
been retrieved as combinations of expectations and variances of suitable
mode quadratures, obtained by combining the entangled modes by linear
optics. The quantities of interest have been obtained tomographically,
processing the whole data set and thus reducing statistical
fluctuations.  Upon exploiting a general model allowing local squeezing
and polarization cross-talking inside the crystal, we have very
precisely described the experimental CM with the theory underlying
parametric downconversion, thus providing a full explanation of
experimental findings.  The reconstructed state is a Gaussian entangled
state close to a two-mode squeezed thermal state, the corresponding
entanglement and nonclassical photon number correlations have been
demonstrated.  We conclude that single homodyne schemes are convenient
and robust setups for the full characterization of OPO signals and, in
turn, represent a relevant tool for quantum technology based on CV
entanglement, e.g., the full characterization of CV Gaussian
channels by input-output signals' characterization. Finally,
making use of a single OPO and a single homodyne detector, our setup
represents also a compact and robust tool for entanglement generation and
characterization.
%%%%%%%%%%%%%%%%%%%%%%%%
\par {\em Acknowledgments}---This work has been partially
supported by CNR-CNISM. MGAP thanks M.~Bondani and A.~Allevi
for discussions.
%%%%%%%%%%%%%%%%%%%%%%%%

%%%%%%%%%%%%%%%%%%%%%%%%

\begin{thebibliography}{99}
\bibitem{pars} The other parameters are given by $\xi_1 \simeq 0.07 e^{i
0.12 \pi}$, $\xi_2 \simeq 0.12 e^{-i 0.12\pi}$, $\beta \simeq 0.13\pi
e^{-i 0.17 \pi}$.
\end{thebibliography}
\end{document}